\documentclass[a4paper]{jpconf}
\usepackage{graphicx}
\usepackage{graphics}
\usepackage{iopams}
\usepackage{amsfonts}

\begin{document}
\newcommand{\half}{\mbox{$\textstyle \frac{1}{2}$}}
\newcommand{\quat}{\mbox{$\textstyle \frac{1}{4}$}}
\newcommand{\octa}{\mbox{$\textstyle \frac{1}{8}$}}
\newcommand{\rd}{{\rm d}}
\newcommand{\ri}{{\rm i}}
\newcommand{\re}{{\rm e}}
\title{On quantum microcanonical equilibrium}

\author{Dorje~C.~Brody${}^1$, Daniel~W.~Hook${}^2$, and
Lane~P.~Hughston${}^3$}

\address{${}^1$Department of Mathematics, Imperial College, London
SW7 2BZ, UK}

\address{${}^2$Blackett Laboratory, Imperial College,
London SW7 2BZ, UK}

\address{${}^3$Department of Mathematics,
King's College London, The Strand, London WC2R 2LS, UK}

\begin{abstract}
A quantum microcanonical postulate is proposed as a basis for the
equilibrium properties of small quantum systems. Expressions for the
corresponding density of states are derived, and are used to
establish the existence of phase transitions for finite quantum
systems. A grand microcanonical ensemble is introduced, which can be
used to obtain new rigorous results in quantum statistical
mechanics.
\end{abstract}


\section{Introduction}
\label{sec:1}

The purpose of this paper is to examine properties of quantum
systems in thermal equilibrium. Questions that arise in this
context, for example, are: ``What is the state of a system in
equilibrium?'' or ``What is the temperature of an isolated system in
equilibrium?'' In the case of a classical system immersed in a heat
bath, the equilibrium distribution takes the Gibbs form $\exp(-\beta
H)/Z(\beta)$, where $\beta=1/k_BT$ is the inverse temperature of the
bath. What about a quantum system? Is the equilibrium state given by
the Gibbs density matrix $\exp(-\beta {\hat H})/Z(\beta)$? If so,
how does one verify that the parameter $\beta$ appearing in the
density matrix is the inverse temperature of the bath?

To investigate questions of this kind it is useful to consider first
the classical situation. In the classical case, we take the system
and the bath as a whole and regard this as a single isolated system.
The Hamiltonian (symplectic) structure of classical phase space
$\Gamma$ then allows us to define the density of states
$\Omega(E)=\int_\Gamma\delta(H(x)-E)\rd V$ as the weighted volume of
the phase space occupied by states with energy $E$. In equilibrium
the state of the system maximises entropy and thus is given by a
uniform distribution over the energy surface; this can be derived if
the Hamiltonian evolution exhibits ergodicity. The entropy of the
equilibrium state is thus given by $S(E)=k_B\ln\Omega(E)$, and the
temperature is defined by the thermodynamic relation $T\rd S=\rd E$.
These are the necessary ingredients for the consideration of the
equilibrium properties of a small subsystem. In particular, under a
set of reasonable assumptions, it is possible to deduce, by use of
the law of large numbers, that the equilibrium properties of a small
subsystem are described by the Gibbs state. A complete derivation of
these results is outlined in the seminal work of
Khinchin~\cite{khinchin1}. Although the derivation of the
equilibrium state is surprisingly complicated, once the relevant
assumptions are specified, there are no ambiguities in the matter,
and familiar results associated with the canonical ensemble can be
obtained rigorously.

The situation is markedly different in the case of a quantum system.
First, in the usual Hilbert space formulation of quantum mechanics
it is not clear how one can exploit the Hamiltonian structure. This
leads to a difficulty in defining the temperature of a closed
system. Second, since no rigorous derivation of the temperature
exists (at least for finite quantum systems), it is not possible to
verify whether the parameter $\beta$ appearing in the Gibbs density
matrix agrees with the inverse temperature of the bath.

In the literature on quantum statistics it is often postulated that
the microcanonical density matrix of a quantum system with
eigenenergy $E_i$ is given by the projection operator onto the
Hilbert subspace spanned by states with that energy, normalised by
the dimensionality $n_{E_i}$ of that subspace. The entropy is then
defined by the expression $S=k_B\ln n_{E_i}$. A rigorous derivation
of this density matrix is given by Khinchin~\cite{khinchin2};
however, the assumptions required to obtain the result go beyond
those required for the classical case. In particular, it is
necessary to forbid all superpositions of states with different
energy. The exclusion of general superpositions, however,
contradicts the superposition principle of quantum mechanics. This
incompatibility between quantum mechanics and quantum statistical
mechanics is an issue that has troubled many authors. For example,
Schr\"odinger remarked in this connection that ``. . . this
assumption is irreconcilable with the very foundations of quantum
mechanics'', and that ``. . . to adopt this view is to think along
severely `classical' lines'' \cite{schrodinger}. Confronted with
this apparent contradiction, Schr\"odinger was nonetheless able to
offer an argument to show, in effect, that in thermodynamic limit
(where the number of particles in the system approaches infinity)
the assumption that general superpositions are forbidden is
justified \cite{schrodinger}.

There is another important shortcoming in the familiar derivation of
quantum statistical mechanics, namely, that the entropy is a
discontinuous function of the energy. As a consequence, the
temperature of a finite isolated system is undefined. This issue is
addressed by Griffiths~\cite{griffiths}, who demonstrated the
existence of a thermodynamic limit in which thermodynamic functions
are well defined. We thus see that to make sense of the conventional
approach to quantum statistics, a ``macroscopic'' limit is required.
In this limit, however, we expect quantum systems to behave
semiclassically so that superpositions, in particular, are excluded.
For finite quantum systems, these issues remain unresolved.

While the notion of a thermodynamic limit was justified both
theoretically and experimentally some forty years ago, there have
been experiments carried out on quantum systems over the past decade
that involve small numbers of particles (see Gross~\cite{gross} and
references cited therein). In particular, phase transitions have
been observed in small systems---for example, the spherically
symmetric cluster of $139$ sodium atoms exhibits a solid-to-liquid
phase transition at about $267$~K~\cite{schmidt}. Such experiments
demonstrate the breakdown of the conventional approach in which
phase transitions are predicted only in thermodynamic limits.

To obtain an equilibrium distribution that is well defined for
finite systems, and to address the issue of the observed finite-size
phase transitions, we have recently introduced an alternative
formulation to quantum microcanonical equilibrium \cite{bhh}. The
idea is to follow the derivation of the traditional result, as
outlined in Khinchin~\cite{khinchin2}, as closely as possible, but
to relax just one of the assumptions; namely, for a fixed energy
$E$, we allow the system to be in a superposition of energy
eigenstates with distinct eigenvalues.

\section{Thermodynamic equilibrium}
\label{sec:2}

The idea of the new microcanonical equilibrium can be described
heuristically as follows. We consider a gas consisting of a large
number $N$ of weakly-interacting identical quantum molecules. As in
the conventional approach, the intermolecular interactions are
assumed strong enough to allow the gas to thermalise but weak enough
so that, to a good approximation, the total system energy can be
written as $\sum_{i=1}^N {\hat H}_i\approx{\hat H}_{\rm total}$,
where $\{{\hat H}_i\}_{i=1,2,\ldots,N}$ are the Hamiltonians of the
individual constituents. If the composite system is in isolation,
then the total energy is a fixed constant: $\sum_{i=1}^N
\langle{\hat H}_i \rangle=E_{\rm total}$. Now consider the result of
a hypothetical measurement of the energy of one of the constituents.
In equilibrium, the state of each constituent should be such that
the average outcome of an energy measurement should be the same;
that is, $\langle{\hat H}_i\rangle=E$, where $E=N^{-1}E_{\rm
total}$. In other words, the equilibrium state of each constituent
must lie on the energy surface ${\mathcal
E}_E=\{|\psi\rangle\large|~\langle\psi|{\hat H}_i|\psi\rangle=E\}$
in the pure-state manifold for that constituent. Since $N$ is large,
this will ensure that the uncertainty in the total energy of the
composite system, as a fraction of the expectation of the total
energy, is vanishingly small.

It is convenient to describe the distribution of the various
constituent pure states, on their respective energy surfaces, as if
we were considering a probability measure on the energy surface
${\mathcal E}_E$ of a single constituent. In reality, we have a
large number of approximately independent constituents; but owing to
the fact that the respective state spaces are isomorphic we can
represent the behaviour of the aggregate system with the
specification of a probability distribution on the energy surface of
a single ``representative'' constituent.

In thermal equilibrium the resulting distribution should be uniform
over the energy surface ${\mathcal E}_E$ since it must maximise the
entropy. Therefore, the density of states is given by
\begin{eqnarray}
\Omega(E) = \int_\Gamma \delta(H(\psi)-E)\rd V_\Gamma.
\end{eqnarray}
Here, $\Gamma$ denotes the pure state manifold and $\rd V_\Gamma$ is
the associated Fubini-Study volume element of $\Gamma$. Once
$\Omega(E)$ is specified, the entropy is given by $S(E)=k_B\ln
\Omega(E)$. It follows that the temperature and the specific heat
can be deduced from thermodynamic relations $T\rd S=\rd E$ and
$C(T)=\rd E/\rd T$. A short calculation shows that
\begin{eqnarray}
k_BT=\frac{\Omega(E)}{\Omega'(E)}, \quad {\rm and}\quad C(T) =
\frac{k_B(\Omega')^2}{(\Omega')^2-\Omega\Omega''}. \label{eq:2}
\end{eqnarray}

The advantage of the present formulation over the traditional
approach is that the entropy is a continuous function of the energy.
As a consequence, thermodynamic functions such as those in
(\ref{eq:2}) are well defined for finite quantum systems. However,
to justify the term ``temperature'' for the ratio $\Omega/\Omega'$
we must show its properties are consistent with the requirements of
thermodynamic equilibrium. For this purpose, consider two
independent systems, each in equilibrium, with state densities
$[\Omega_1(E_1)]^{N_1}$ and $[\Omega_2(E_2)]^{N_2}$. We let them
interact for a period of time, during which energy $\epsilon$ is
exchanged. We then separate them and let them relax again to
equilibrium. Because of the interaction the state densities of the
systems are now $[\Omega_1 (E_1 + \epsilon/N_1)]^{N_1}$ and
$[\Omega_2(E_2- \epsilon/ N_2)]^{N_2}$. The value of $\epsilon$ is
determined so that the total entropy $S(E)=k_B\ln [\Omega_1(E_1 +
\epsilon/N_1)]^{N_1} [\Omega_2(E_2- \epsilon/N_2)]^{N_2}$ is
maximised. This condition is satisfied if and only if $\epsilon$ is
such that the temperatures of the two systems defined according to
(\ref{eq:2}) are equal. It follows that our definitions are
thermodynamically consistent.

\section{Expressions for the density of states}
\label{sec:3}

Let us now try to obtain a direct representation for the density
of states $\Omega(E)$ in terms of the energy eigenvalues. We
consider first the two-level system with energy eigenvalues
$E_1,E_2$. The Fubini-Study volume element for the pure state
manifold is $\rd V_\Gamma= \frac{1}{4}\sin\theta\rd\theta\rd\phi$,
where $0\leq\theta\leq\pi$ and $0\leq\phi<2\pi$. Since the energy
expectation in a generic state $|\psi\rangle =
\cos\half\,\theta|E_2\rangle+ \sin\half\, \theta\,\re^{{\rm
i}\phi}|E_2\rangle$ is $E_2\cos^2\half\,
\theta+E_1\sin^2\half\,\theta=\half(E_2-E_1)(1+\cos\theta)+E_1$,
the density of states is
\begin{eqnarray}
\Omega(E)= \frac{1}{8\pi} \int_{-\infty}^\infty \rd \lambda
\int_0^{2\pi}\rd \phi \int_0^{\pi}\rd\theta \,\re^{-{\rm i}
\lambda\left({\bar E}(1+\cos\theta)+E_1-E\right)} \sin\theta.
\label{eq:4}
\end{eqnarray}
Here, we have made use of the integral representation for the
delta-function, and we have also defined ${\bar E}=(E_2-E_1)/2$. By
use of the relation $\int_{-\infty}^\infty\rd\lambda\, \re^{-{\rm i}
b\lambda} \lambda^{-1} \sin(a\lambda)=\pi$ for $a>b$ and $=0$ for
$a<b$ we thus deduce that $\Omega(E)=\pi/(E_2-E_1)$ for $E_1\leq
E\leq E_2$, and $\Omega(E)=0$ otherwise.

An analogous calculation can be performed for a three level system.
If we let $E_1,E_2,E_3$ denote the eigenvalues of the Hamiltonian,
then the energy constraint for a generic state $|\psi\rangle =
\sin\half\,\theta\cos\half\,\varphi|E_3\rangle+ \sin\half\,
\theta\sin\half\,\varphi\,\re^{{\rm i}\xi}|E_2\rangle +
\cos\half\,\theta\,\re^{{\rm i}\eta}|E_1\rangle$ is given by
$E_3\sin^2\half\,\theta\cos^2\half\,\varphi + E_2 \sin^2\half\,
\theta\sin^2\half\,\varphi + E_1\cos^2\half\,\theta=E$. Since the
Fubini-Study volume element in this case is $\rd V_\Gamma =
\frac{1}{32} \sin\theta(1-\cos\theta)\sin\varphi \rd\theta \rd
\varphi \rd\xi\rd\eta$, we carry out the relevant integration and
obtain
\begin{eqnarray}
\Omega(E)=\frac{\pi^2(E-E_1)}{(E_3-E_1)(E_2-E_1)} \quad {\rm or}
\quad \Omega(E) = -\frac{\pi^2(E-E_3)}{(E_3-E_1)(E_3-E_2)},
\label{eq:6}
\end{eqnarray}
depending on $E_1\leq E\leq E_2$ or $E_2< E\leq E_3$.

\begin{figure}[h]
\includegraphics[width=20pc,height=10.0pc]{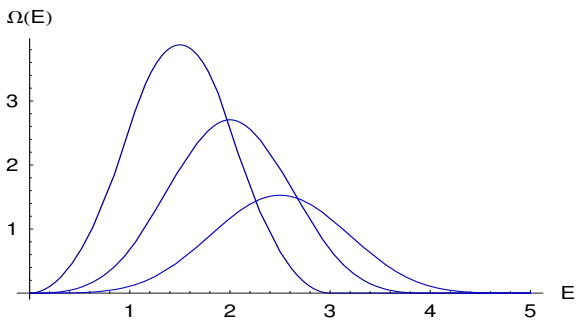}\hspace{2pc}%
\begin{minipage}[b]{16pc}
\caption{\label{fig:1}Density of states $\Omega(E)$ given in
(\ref{eq:7}) associated with nondegenerate $(n+1)$-level systems for
$n=3,4,5$, with energy eigenvalues $E_k=k$ $(k=0,\ldots,n)$. For an
$(n+1)$-level system, $\Omega(E)$ is a piecewise polynomial of
degree $n-1$, and is $n-2$ times differentiable in energy $E$.}
\end{minipage}
\end{figure}

By pursuit of this line of argument we deduce more generally that
the density of states $\Omega(E)$ is given by a piecewise polynomial
function of energy $E$. In particular, if the energy spectrum is
nondegenerate, then we have the representation
\begin{eqnarray}
\Omega(E)=\frac{(-\pi)^n}{(n-1)!} \sum_{k=1}^{n+1} (E_k-E)^{n-1}
\prod_{l\neq k}^{n+1} \frac{{\mathbf 1}_{\{E_k> E\}}}{E_l-E_k}.
\label{eq:7}
\end{eqnarray}
Here ${\mathbf 1}_{\{A\}}$ denotes the indicator function (${\mathbf
1}_{\{A\}}=1$ if $A$ is true, and $0$ otherwise). To offer an
intuition for the behaviour of the density of states, examples of
$\Omega(E)$ are shown in Figure~\ref{fig:1}.

Once the density of states $\Omega(E)$ is obtained for the
microcanonical equilibrium, thermal expectation values of physical
observables in the corresponding canonical distribution can be
computed by use of the canonical partition function $Z(\beta)$,
which is the Laplace transform of $\Omega(E)$. When energy
eigenvalues are nondegenerate, we have
\begin{eqnarray}
Z(\beta) = \sum_{k=1}^{n+1}\re^{-\beta E_{k}} \prod_{l=1,\neq
k}^{n+1}\frac{\pi}{\beta(E_{l}-E_{k})} . \label{eq:9}
\end{eqnarray}
A line of argument in Khinchin~\cite{khinchin1} for classical
systems can then be applied here in the quantum context to prove
that the parameter $\beta$ appearing in (\ref{eq:9}) for the
canonical partition function agrees with the microcanonical
definition of temperature in (\ref{eq:2}).

\section{Quantum phase transitions}
\label{sec:4}

An interesting consequence of the microcanonical framework, whether
classical or quantum, is that the density of states in general need
not be an analytic function for finite systems. In contrast, the
partition function in the canonical counterpart is necessarily
analytic. In other words, while it is necessary in the canonical
framework to take thermodynamic limit to describe phase transitions,
in the microcanonical formalism this is not the case. Therefore, an
approach based on microcanonical equilibrium might provide an
adequate description of the phase transitions for small systems
observed in the laboratory. We note in this connection that there
are many classical systems for which finite-size phase transitions
are predicted in microcanonical equilibrium~\cite{kastner,pettini}.

\begin{figure}[h]
\begin{center}
\includegraphics[width=0.76\textwidth]{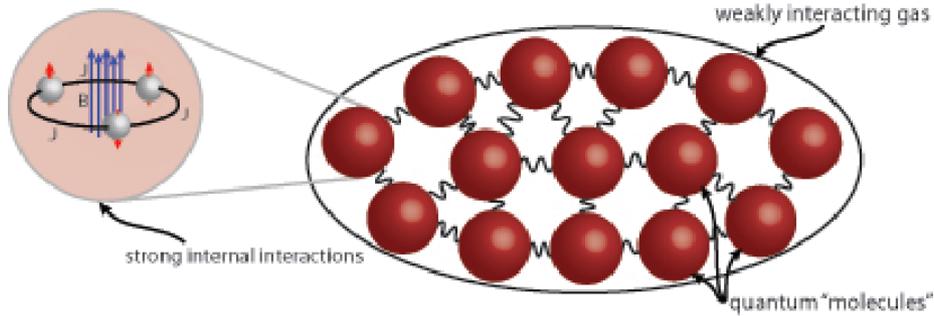}
\caption{\label{fig:2} A gas of quantum Ising chains. The gas
consists of a large number of weakly interacting quantum
molecules. Each molecule is modelled by a quantum Ising chain of
three strongly interacting spin-$\half$ particles. }
\end{center}
\end{figure}

In quantum microcanonical equilibrium, the breakdown of analyticity
of $\Omega(E)$ gives rise to phase transitions in the sense that
discontinuities in the higher-order derivatives of $\Omega(E)$
emerge. Specifically, if we solve the first equation in (\ref{eq:2})
for the energy to obtain $E(T)$, then for a system with $n+1$
nondegenerate energy eigenvalues, the $(n-1)$-th derivative of the
energy with respect to the temperature has a discontinuity.

\begin{figure}[h]
\includegraphics[width=20pc,height=12.5pc]{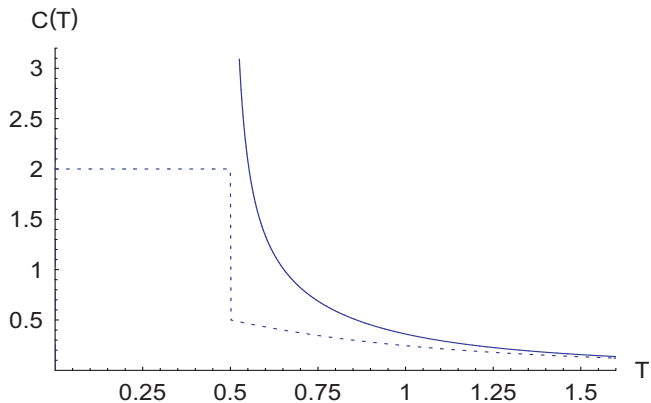}\hspace{2pc}%
\begin{minipage}[b]{16pc}
\caption{\label{fig:3}Specific heat for a nondegenerate four-level
system (dotted line, $n=3$, $E_j=0,1,2,3$), and a quantum Ising
chain having a four distinct degenerate eigenvalues (solid line,
$J=1/4$, $B=1$). The plots illustrate the existence of second-order
phase transitions. In the quantum Ising chain, we have $C(T)\sim
(T-T_c)^{-2}$ away from $T_c$, whereas in the vicinity of $T_c$ we
have $C(T)\sim (T-T_c)^{-13}$ for $T>T_c$.}
\end{minipage}
\end{figure}

As an illustration we consider the specific heat $C(T)$ for a gas of
weakly interacting molecules, where each molecule is modelled by a
strongly interacting chain of three Ising-type spins (see
Figure~\ref{fig:2}). The molecular Hamiltonian is ${\hat H} = -J
\sum_{k=1}^3 \sigma_z^k \sigma_z^{k+1} - B \sum_{k=1}^3 \sigma_z^k$,
where $\sigma_z^k$ is the third Pauli matrix for spin $k$, and $J,B$
are constants. For this system, the specific heat grows rapidly in
the vicinity of the critical point $T_c=(2J+B)/3k_B$, where the
system exhibits a discontinuity in the second derivative of the
specific heat. The plot of the specific heat is shown in
Figure~\ref{fig:3}, along with the corresponding plot for a simple
four-level molecular gas; the latter exhibits a second-order phase
transition at the critical temperature $k_BT_c = \half \varepsilon$
and critical energy $E_c = \varepsilon$, where $\varepsilon$ is the
spacing of energy eigenvalues.

\section{Towards quantum grand microcanonical equilibrium}
\label{sec:5}

In the foregoing discussion we have made use of the energy
conservation property of the unitary evolution to introduce a
\emph{quantum microcanonical hypothesis} which asserts that in
equilibrium, every quantum state with given energy $E$ is realised
with an equal probability. This hypothesis can be refined in the
following manner, leading to what might appropriately be called the
quantum \textit{grand microcanonical hypothesis}.

For a given quantum mechanical system there are $n$ linearly
independent conserved observables, where $n+1$ is the Hilbert space
dimensionality. Therefore, when an isolated quantum system with a
generic Hamiltonian evolves unitarily, the associated dynamics
exhibit ergodicity on the toroidal subspace ${\mathcal T}^n
\subset{\mathcal E}_E$ of the energy surface determined by
simultaneously fixing the expectation values of the commuting family
of observables (cf.~\cite{brody}). A theorem of
Birkhoff~\cite{khinchin1} applies to show that the dynamical average
of an observable can be replaced by the ensemble average with
respect to a uniform distribution over ${\mathcal T}^n$. The density
of states is then determined by the weighted volume of the subspace
${\mathcal T}^n$ of the quantum state space.

In the case of the energy observable, the conjugate variable is
given by the inverse temperature. For other observables belonging to
the commuting family, the associated conjugate variables can be
thought of as \emph{generalised chemical potentials}. In this
respect, a refinement of the microcanonical postulate leads to an
ensemble of the grand canonical form.

Let us consider the simplest nontrivial example $n=2$. We choose the
two projection operators ${\hat\Pi}_1=|E_1\rangle\langle E_1|$ and
${\hat\Pi}_2= |E_2\rangle\langle E_2|$ for the independent pair of
commuting observables. Since these observables are conserved, we let
the two constraints be $\langle{\hat\Pi}_1\rangle=p$ and $\langle
{\hat\Pi}_2\rangle=q$. It follows from the resolution of identity
that $\langle {\hat\Pi}_3\rangle=1-p-q$. In terms of the usual
parametrisation, these constraints read $\cos^2\half\,\theta=p$ and
$\sin^2\half\,\theta \sin^2\half\,\varphi=q$, respectively. The
generalised density of states $\Omega(p,q)=\int_\Gamma
\delta(\langle{\hat\Pi}_1\rangle-p)
\delta(\langle{\hat\Pi}_2\rangle-q) \rd V_\Gamma$ can then be
calculated to yield
\begin{eqnarray}
\Omega(p,q)=\quat\pi^2\left(\Upsilon(p)-\Upsilon(p+q-1) \right)
\left(\Upsilon(q)-\Upsilon(q-1) \right), \label{eq:10}
\end{eqnarray}
where $\Upsilon(x)=-1$ for $x\leq0$ and $\Upsilon(x)=1$ for $x>0$.
We have $\Omega(p,q)=\pi^2$ in the range $0\leq p,q\leq1$ and $0\leq
1-p-q \leq1$. To establish its relation with the density of states
$\Omega(E)$ we solve the energy constraint $pE_1+qE_2+(1-p-q)E_3=E$
for, say, $p$, then substitute the result in $\Omega(p,q)$, and
integrate over $q$ from $0$ to $1$. The temperature of the system
can then be obtained by differentiation. It would be of interest to
further investigate properties of the grand microcanonical
equilibrium for general systems, which in our view holds the promise
for many new rigorous results in quantum statistical mechanics.

\ack DCB acknowledges support from The Royal Society. DWH thanks the
organisers of the DICE2006 conference in Piombino, Italy, 11-15
September 2006 where this work was presented. The authors thank
M.~Parry for comments.

\end{document}